\documentclass{article}

\usepackage{arxiv}

\usepackage[utf8]{inputenc} 
\usepackage[T1]{fontenc}    
\usepackage{hyperref}       
\usepackage{url}            
\usepackage{booktabs}       
\usepackage{amsfonts}       
\usepackage{nicefrac}       
\usepackage{microtype}      
\usepackage{lipsum}
\usepackage{graphicx}
\usepackage{amsmath}
\usepackage{float}
\graphicspath{ {./images/} }

\title{Advantages of OKID-ERA Identification in Control Systems. An Application to the Tennessee Eastman Plant}

\author{
 Sergio F. Yapur \\
  Facultad de Ingenier\'ia Qu\'imica \\
  Universidad Nacional del Litoral\\
  Santiago del Estero 2829 (3000) Santa Fe \\
  \texttt{syapur@fiq.unl.edu.ar} \\
}

\begin{document}
\maketitle
\begin{abstract}
Data-driven OKID-ERA identification of the open-loop Tennessee Eastman plant is performed to obtain a linear model for control design purposes. Analysis such as numerical conditioning, output response errors, and zero-pole mappings highlight some definite advantages of the OKID-ERA approach when compared with models derived from typical linearization techniques. The plant under study is a recognized benchmark in the field of plant-wide control systems.

\end{abstract}


\section{Introduction}

Process simulation has become commonplace in industries with complex processes. Its importance grows further thanks to trends like IIoT or Industry 4.0. In the field of control systems, linear models are particularly useful. They usually result from a linearization around an operative point of the plant. In turn, linear models offer a great deal of information to produce a control system solution. A so-called solution includes structural decisions like input-output pairing, as well as controller-type selection and tuning methods. Modeling from linearization techniques have prove their usefulness extensively. Nevertheless, in some cases, this kind of models can exhibit a variety of issues that might prevent an effective final control scheme on the plant.

Certain nonlinear features result in modeling issues. For instance, a given process can have fast dynamic associated with gas pressure changes while exhibiting low dynamics in heat transfer through solid phase. In extreme cases, this large dynamic range leads to a ill-conditioned system matrix on the linearization model. This situation may render unfruitful a number of control-system design techniques.

Another source of errors comes from high-dimensional systems. Complex plants usually present hundreds of input-output variables, together with a similar amount of internal states. As the size of the system increases, it becomes increasingly difficult for the control system to keep the plant around a fixed operative point. Actually, it may not be desirable to reach such a tight control for large plants, for it represents higher costs of control system installation, operation and maintenance. Needless to say, a strict control of critical variables is of upmost importance due to economic and safety reasons. However, extending this criterion to non-critical variables may be detrimental. In any case, linearization-based models best represents the process around the associated operative point. Therefore, deviations due to non-controlled variables, such as perturbations, noises or nonlinear variable couplings, detach the model from the actual process, possibly affecting control decisions. Considerable care must be exercised with the concept of operative-point, especially concerning complex plants. 

Additionally, it is not rare to find different sampling rates across signals coexisting in the same chemical process. A major cause of this situation is the relatively long processing times of chemical concentrations compared to other variables like temperature or pressure. Consequently, discrete and continuous-like signals coexist in the same plant. Recalling that linearization methods generally use numerical approximations of partial derivatives, any implementation on a black-box system with mixed-sampling should proceed with caution. Different notions of proximity should be used for variables with dissimilar sampling rates, in order to avoid significant approximation errors. However, this is not usually the case and most numerical approximations of the Jacobian matrix use the same time-step anywhere. As a possible way to solve this issue, a subset of signals should be processed independently for every given time-step.

Yet another source of errors arise from the approximation of dead-times within the plant variables. Most real processes present this phenomenon to some extend. Even the location of sensors in the piping is critical to avoid unnecessary large dead times. The well-known issue of dead times in control systems is particularly harmful regarding synthesis of linear models from data alone. Many identification algorithms introduces high order terms to approximate dead times. This approach, however, modifies the identified dynamics regarding the real process. On the other hand, some of the methods to assess dead-times from data do not work properly when applied to noisy, MIMO systems. Presumably, one of the best method might be the Information Theoretic Delay Criterion due to the general nature of its theoretical foundation \cite{YapurTesisDI}.

This article is organized as follows. The next section describes the modeling approach, including the selection of algorithms featured. The third section presents the study case, a well-known test problem in the plant-wide control community. Then, the result section highlights the main attributes of this approach, as well as comparisons with more typical modeling techniques. Finally, conclusions are drawn in the final section.


\section{Identification Approach}

Dynamical system theory is currently expanding to data-driven approaches by means of machine learning and big-data. In this context, promising system identification techniques emerge naturally from black-box analysis where input-output mappings are obtained from training data.

As mentioned in the Introduction, the relevance of a fixed operative point decreases as the plant size increases. Consequently, one may expect that a system-identification approach provides a linear model that better captures the overall behavior of the plant, since it does not depend necessarily on a particular operative point. Instead, the model relies on a region of operation associated with the available data.

Among the many existing identification techniques \cite{Tangirala2015, Ljung1999}, the Eigenvalue Realization Algorithm (ERA) enhanced with the Observer-Kalman Identification (OKID) seems to be a suitable choice for the Tennessee Eastman plant. This is partly because only data with noise is available, thus requiring a robust identification method. Also, the strong non-linearity of the plant suggest that signals may rapidly diverge from the initial point, so several variables rapidly spread over a region in space.

\subsection{Eigensystem Identification Algorithm}

ERA is based on the minimal realization theory of Ho and Kalman \cite{Ho_Kalman}. It was introduced in 1985, but it has gained renewed relevance in the last years with the consolidation of data-driven engineering. It has a straightforward implementation, making this method widely used. Moreover, it is suitable to obtain a low-order, balanced linear model directly from impulse response data. Results from this technique were shown to be equivalent to Balanced Proper Orthogonal Decomposition (BPOD) \cite{Ma2011}. 

This identification technique produces models from data alone, without the need of computationally expensive adjoint simulations. It also has profound links to Dynamic Mode Decomposition (DMD), since they are algorithmically similar for impulse response input-data. 

Given measurements of impulse response, ERA obtains the parameters of a linear, discrete-time dynamical system with actuation $\mathbf{u}$, outputs $\mathbf{y}$, and internal states $\mathbf{x}$

\begin{eqnarray}
	\mathbf{x}_{k+1} &=& \mathbf{A x}_k + \mathbf{B u}_k \label{eq:linModel1} \\
	\mathbf{y}_k &=& \mathbf{C x}_k  \label{eq:linModel2}
\end{eqnarray}

An discrete-time impulse response is required for this algorithm to work. This is defined as

\begin{equation}
    \mathbf{u}_k = 
    \left\{
    \begin{array}{ll}
         \mathbf{I} & \mbox{for $k=0$} \\
         \mathbf{0} & \mbox{for $k \in \mathbb{N}$}.
    \end{array}
    \right.
\end{equation}

Output measurements result from

\begin{equation}
    \label{eq:yvalues}
    \mathbf{y}_k^\delta = 
    \left\{
    \begin{array}{ll}
         \mathbf{I} & \mbox{for $k=0$} \\
         \mathbf{C A}^{k-1} \mathbf{B} & \mbox{for $k \in \mathbb{N}$}.
    \end{array}
    \right.
\end{equation}

These values are arranged in Hankel matrices, which are highly-redundant data structures. Redundancy is central to make the identification robust, meaning to better capture the dynamic modes of the system. Mathematically, a Hankel matrix results from $\mathbf{H} = \mathcal{O}_d \mathcal{C}_d$, where $\mathcal{C}_d$ and $\mathcal{O}_d$ are the discrete-time controllability and observability matrices respectively \cite{Brunton2019}. However, these matrices are not required for ERA to work, since the elements of $\mathbf{H}$ arise directly from the observations. To visualize this, note that the Hankel matrix $\mathbf{H}$ and its time-shifted version $\mathbf{H'}$ can be rewritten as 

\begin{eqnarray}
    \label{eq:Hankel}
    \mathbf{H}_{ij} &=& \mathbf{C A}^{i+j-2} \mathbf{B} \quad \mbox{$i=1,...,s$, \quad $j=1,...,m-s$} \\    
    \label{eq:HankelShift}
    \mathbf{H'}_{ij} &=& \mathbf{C A}^{i+j-1} \mathbf{B} \quad \mbox{ $i=1,...,s$, \quad $j=1,...,m-s$}.
\end{eqnarray}

The elements of these arrays are actually observations as shown in Ec. \ref{eq:yvalues}. They are also known as Markov parameters, since there is an important connection between Hankel matrices and Hidden Markov Models (HMM) \cite{Kutz}.

A reduced Singular Value Decomposition (SVD) of the form $\mathbf{H} \approx \mathbf{U}_r \mathbf{\Sigma}_r \mathbf{V}_r$, with $r$ denoting a truncation index, reveals the dominant temporal patterns of the system. Also, using SVD on both $\mathbf{H}$ and $\mathbf{H'}$, enables a low-dimensional approximation of the system matrices $\mathbf{A, B, C, D}$. This means that the dominant dynamics not only is reconstructed from data alone, but also in a computationally efficient way by using either economy or truncation SVD \cite{Kutz}. 

A distinctive feature of ERA is that it produces a balanced realization of the system provided that enough data is collected. Otherwise, it will only approximately balance the system. This is specially useful for control purposes.

\subsection{Observer Kalman Filter Identification}

Despite its advantages, ERA has evident limitations. Firstly, it requires impulse response data to work, which may be unrealistic or difficult to obtain in practice. It is also sensitive to noise, meaning it degrades the quality of the model. Moreover, a large separation of timescales in the system requires a vast amount of data.

OKID addresses the ERA drawbacks by processing raw data from arbitrary input with noise. The resulting output is denoised, linear impulse response data. In other words, the method provides the Hankel elements $\mathbf{H}_{ij}$ (or Markov parameters) from any given input-output data, without explicitly calculating the system matrices. In this manner, OKID becomes a powerful pre-processing stage for ERA. 
 
Both ERA and OKID were developed to construct input-output state-space models for aerospace applications, where system rank tend to be higher than the number of sensor measurements. Anyhow, it is straightforward obtaining a low rank model from these formulations. In principle, this is as simple as choosing a truncation value $r$ in the SVD of $\mathbf{H}$. Although the selection is not obvious in presence of noise and disturbances \cite{Tangirala2015}, heuristic rules are available to help finding an appropriate value of $r$.

Instead of the original discrete-time system, the OKID method adopts an optimal observer system

\begin{eqnarray}
    \hat{\mathbf{x}}_{k+1} &=& \mathbf{A} \hat{\mathbf{x}}_k + \mathbf{K}(\mathbf{y}_k - \hat{\mathbf{y}}_k) + \mathbf{Bu}_k \\
    \hat{\mathbf{y}}_k &=& \mathbf{C\hat{x}}_k + \mathbf{Du}_k
\end{eqnarray}

By combining the former expressions, the state $\mathbf{\hat{x}}_k$ is estimated as a function of an extended vector $[\mathbf{u}_k, \mathbf{y}_k]^T$

\begin{eqnarray}
    \label{eq:obsKalman}
    \hat{\mathbf{x}}_{k+1} &=& \mathbf{\Bar{A}} \hat{\mathbf{x}}_k + \left[\mathbf{\Bar{B}},\mathbf{K}\right]
    \left[ 
    \begin{array}{c}
         \mathbf{u}_k \\
         \mathbf{y}_k 
    \end{array}
    \right], \quad \text{where} \\
    \mathbf{\Bar{A}} &=& \mathbf{A-K C}, \quad \text{and} \\
    \mathbf{\Bar{B}} &=& \mathbf{B-K D}.
\end{eqnarray}

For convenience, let us call $\mathbf{\hat{y}}_i^\delta = \mathbf{C \Bar{A}}^i \mathbf{\Bar{B}}$ the observer Markov parameters. The introduction of the observer given in Ec. \ref{eq:obsKalman} acts as an artifice to compress the data and improve system identification results in practice. More precisely, if the system is observable, the Kalman filter $\mathbf{K}$ allows setting optimal pole locations to produce $\mathbf{\hat{y}}_i^\delta \approx 0$ for $i>p$, where $p$ is a sufficiently large integer. The pole locations of $\mathbf{K}$ depend on the noise level and model uncertainty \cite{Juang,Billingsley}.  Notice that data is compressed in the sense that most of the $\mathbf{\hat{y}}_k^\delta$ are negligible. In other words, the observer has the effect of increasing the damping of the system, so fewer observer Markov parameters can describe the (observer) system impulse response. Consequently, a computationally cheaper, truncated system is solved. Finally, a simple iterative relationship allows the reconstruction of the Markov parameters $\mathbf{y}_k^\delta$ from their observer version $\mathbf{\hat{y}}_k^\delta$ \cite{Brunton2019}.


\section{The Tennessee Eastman Plant}

The problem was established as a benchmark for plant-wide control studies almost 30 years ago, although it remains challenging up to date. The original article of Downs and Vogel \cite{DownsVogel} presents a complete description of the process. Only a concise overview is provided in this section.

\begin{figure}[ht]
  \centering
  \makebox[\textwidth][c]{\includegraphics[width=0.9\textwidth]{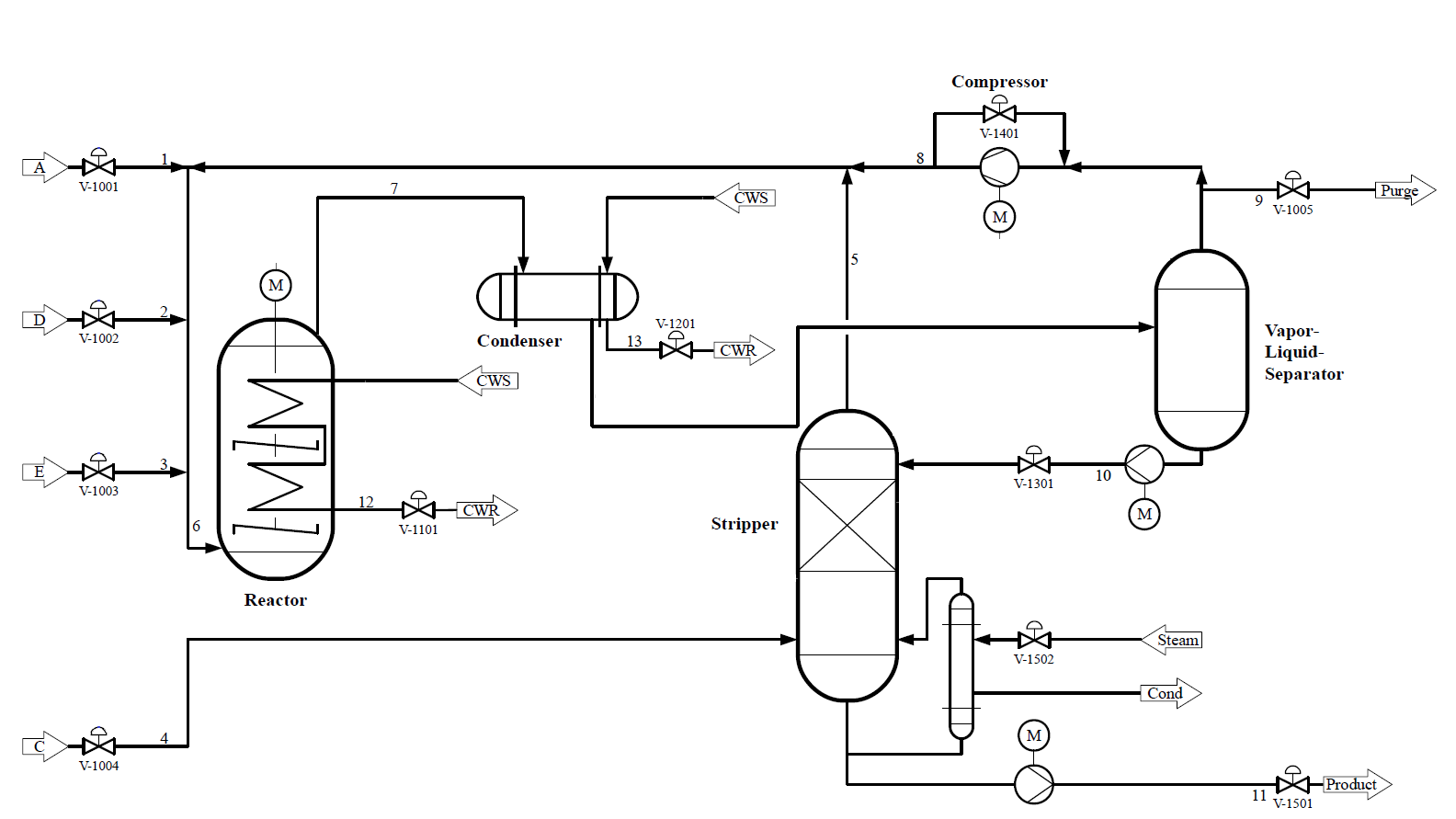}}
  \caption{The Tennessee Eastman process}
  \label{fig:TEplant}
\end{figure}

The Fig. \ref{fig:TEplant} outlines the process flow-sheet, with the inlet and outlet streams on the left and on the right respectively. It represents a petrochemical plant belonging to greater processing complex. From a physical chemical point of view, it is basically a chemical reactor coupled with a flash separator and a stripper. Four chemical reactions occurs simultaneously within an heterogeneous gas-liquid catalytic reactor. They are exothermic and irreversible. The production yield is increased by recycling unreacted reagents to the reactor. This recycle stream increases the interactions between process variables. 

Mathematically, the plant is a strongly nonlinear and open-loop unstable system. Actually, as reported in a previous work \cite{Yapur2021}, it presents chaotic behaviour, meaning that the outcome is unpredictable even when the supporting equations are deterministic. The system has 40 input signals, 12 of which are possible manipulated variables, while the remainder signals are pre-programmed disturbances. Also, there are 41 output signals, 22 of which have continuous measurements, 14 signals have 0.1 h of sampling rate, and 5 signals are sampled every 0.25 h. As described earlier, this mixed sampling may pose numerical issues.

\section{Results}
\label{sec:Results}

The present section retrieve results from linear models previously studied \cite{Yapur2022}, and compare them with the resulting model from the OKID-ERA processing. These previous models were derived from typical linearization procedures around the operative point of the plant. Both are continuous-time linear models, henceforth referred as Lin1 and Lin2, derived from Matlab routines \texttt{linmod} and \texttt{linmodv5} respectively. These procedures constitute typical approaches in the control system community.


\subsection{Modeling Errors}

\begin{figure}[ht]
  \centering
  \makebox[\textwidth][c]{\includegraphics[width=0.65\textwidth]{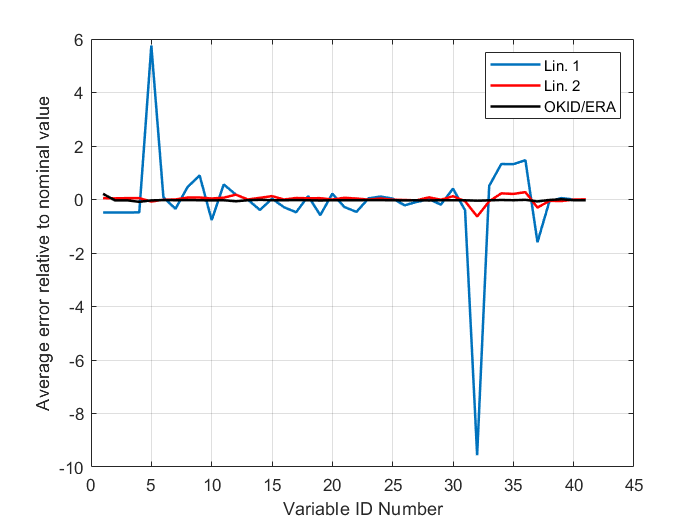}}
  \caption{Time-average deviations with respect nominal values}
  \label{fig:avgErrors}
\end{figure}

Fig. \ref{fig:avgErrors} shows average deviations for the 41 output variables. These are obtained from simulation data of the linear models and the original nonlinear system. To allow a comparison of variables of different units, these deviations are divided by the nominal values, as follows

\begin{equation}
    \label{eq:relErrors}
    e_j = \frac{ \overline{y_j^l(t_i) - y_j^{nl}(t_i)}}{y_j^{nom} }. 
\end{equation}

Note that the subscript $j$ refers to the number of output variable $j=1,...,41$, while superscripts $l, nl$ stand for linear and nonlinear respectively. Also, $t_i$ is a particular value of the discrete time, $i=1,...,n$, and the over-line denote the mean value over time. 

Regarding average errors, it is clear that the OKID-ERA model outperforms the Lin1 model in a remarkable manner, as Lin1 shows errors of magnitude up to nine times the corresponding nominal value. However, the errors of OKID-ERA are only slightly better when compared with Lin2 errors. 

Another performance metric is the maximum deviation of output variables with respect their nominal values, as follows from expression

\begin{equation}
    \label{eq:maxErrors}
    e_j^{max} = \max_i \left( \frac{ y_j^l(t_i) - y_j^{nl}(t_i)}{y_j^{nom}} \right) . 
\end{equation}

\begin{figure}[ht]
  \centering
  \makebox[\textwidth][c]{\includegraphics[width=0.65\textwidth]{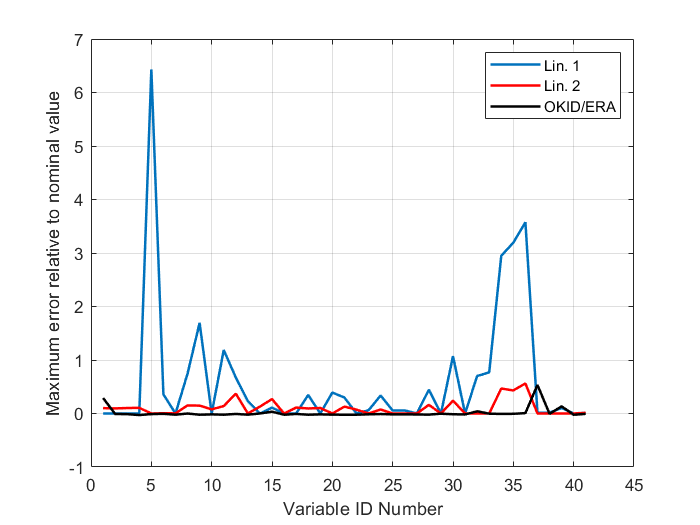}}
  \caption{Maximum deviations with respect nominal values}
  \label{fig:maxErrors}
\end{figure}

These values are shown in Fig. \ref{fig:maxErrors}. In this analysis, the OKID-ERA deviations still remain significantly lower than the Lin1 model. The performance is also noticeable superior than Lin2, which shows greater maximum errors. These conclusions remain valid for OKID-ERA identifications computed over different sets of data from the nonlinear model. Thus, even though it is a data-driven technique applied to a chaotic system, the results are consistent through different realizations. 

Notice that Lin1 model, which is perhaps the default choice in engineering, is in fact the less representative of the actual plant, while OKID-ERA seems to capture the system modes properly only from input-output data.

A possible reason to explain the performance of OKID-ERA shown before is that this method includes data over complete trajectories of the nonlinear responses. In opposition, linearization techniques focus only in a neighborhood of a given point. Because OKID-ERA take advantage of a broader set of data, it can predict nonlinear trajectories more accurately when the system state lies relatively far from the initial point. However, this approach relies strongly on the simulation time length that defines the data. Thus, this factor becomes a tuning parameter in terms of the prediction span of the linear model. 


\subsection{Poles and Zeros}

\begin{figure}[ht]
  \centering
  \makebox[\textwidth][c]{\includegraphics[height=8cm,width=0.6\textwidth]{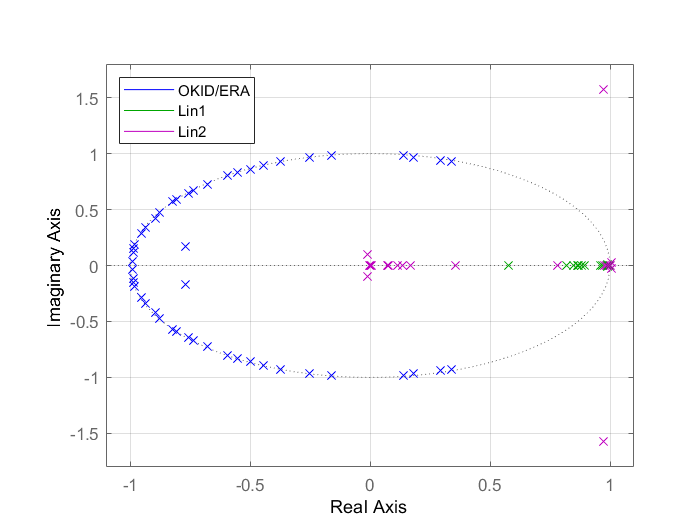}}
  \caption{Poles and Zeros of linear models}
  \label{fig:pzmap}
\end{figure}

The concepts of system poles and transmission zeros are essential to understand the dynamics of MIMO\footnote{Multiple Input, Multiple Output.} systems, and can be found thoroughly along control literature. They are illustrated in Fig. \ref{fig:pzmap} for the three linear models. To make the comparison possible, the continuous model Lin1 was discretized with a  sampling rate of 1 second, which for this type of plants is relatively high frequency. From this graph it is clear that each model has a very distinctive pattern of poles distribution, revealing what in principle may be regarded as different dynamics. Interestingly, the poles of the OKID-ERA model lie on the unit circle, even when the formulation of the associated method is continuous.

What is more, the zero-pole mappings in Fig. \ref{fig:pzmap} suggest different conclusions regarding each model's stability. Even when all three models are unstable, both Lin1 and Lin2 have more poles on the right half-plane, so they have more unstable modes than the OKID-ERA model.



\subsection{Condition Number and Singular Values}

\begin{figure}[ht]
  \centering
  \makebox[\textwidth][c]{\includegraphics[width=1.\textwidth]{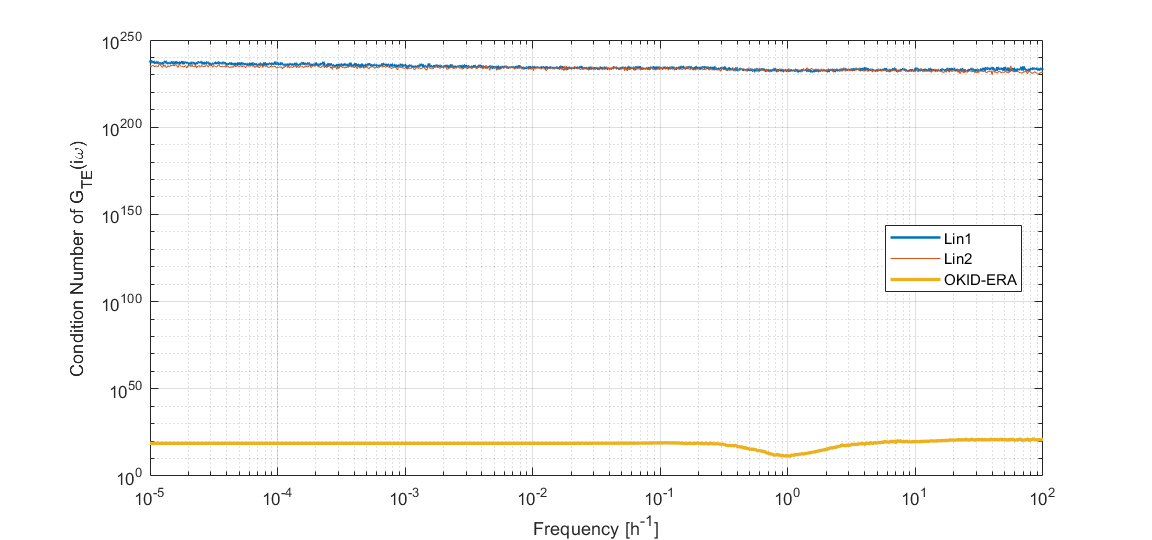}}
  \caption{Condition number vs frequency}
  \label{fig:condNum}
\end{figure}

Matching the nonlinear behaviour of the real system is undoubtedly a desirable feature, but the numerical reliability of a linear model should not be neglected. The evaluation of the condition number over a range of frequencies provides an insight regarding this issue, as presented in Fig. \ref{fig:condNum}. Despite the fact that all linear models are ill-conditioned, the condition numbers of Lin1 and Lin2 are in the order of $10^{230}$, while OKID-ERA shows a number around $10^{20}$. This improvement enables both greater precision and reliability of the model-based calculations.

Naturally, the condition number is a rather broad indicator of numerical issues. Further insight can be obtained from the examination of singular values and loss of rank. However, a detailed analysis of the potential sources of ill-condition is beyond the scope of this work.

\section{Conclusions}

The OKID-ERA algorithm offers the possibility of identifying the model only from data. This data can derive from either experiments or numerical simulations. Also, the input to the system can be arbitrary and the method is robust against signal noise and other uncertainties. These features provide an beneficial framework for complex, real systems. 

The application of the algorithm to the TE plant exhibit promising results. A reduced-order model based on the OKID-ERA method captures the overall behaviour of the nonlinear system remarkably better than traditional linearization-based models, at least in terms of average and maximum relative errors. On the other hand, the analysis of pole-zero mappings suggest distinctive dynamics among the models under study. Finally, the condition number is significantly lower for the OKID-ERA model, which leads to greater numerical stability and precision. These features are especially important for the TE plant due to the chaotic nature of the nonlinear system.



\bibliographystyle{unsrt}  


\end{document}